# CRAW: Combination of Re-keying and Authentication in Wireless Networks for Secure Multicast Increasing Efficiency of Member Join/Leave and Movement

Elina Eidkhani[1], Melisa Hajyvahabzadeh[1],
S. Anahita Mortazavi[1], andAlireza Nemaney Pour[2]

[1]Dept. of IT Engineering, Sharif University of Technology, Kish Island, Iran
elina.eidkhani@gmail.com;melisa.vahabzadeh@gmail.com;mortazavi.anahita
@gmail.com
[2]Dept. of Computer Software Technology Engineering,
Islamic Azad University of Abhar, Iran
pour@abhariau.ac.ir

## ABSTRACT

*Recently, the number of requests for multicast services through the wireless networks has been increased. However, for successful deployment, security and efficiency of content delivery must be provided at first. This paper presents a new approach for secure multicast in wireless networks. This approach, CRAW (Combination of Re-keying and Authentication in Wireless networks) combines member authentication procedure with group key management protocol to provide an efficient group re-keying process. One-time password is proposed for member authentication and CKC (Code for Key Calculation) is suggested for group key management in wireless networks. In fact, the combination of authentication with group key management in wireless networks results in a simple and secure mechanism both for authentication and group key management while mobile members join/leave a group or move inter-area. Simulation results show that CRAW reduces re-keying overhead at join from $O(log_2 n + 1)$ to $O(1)$ while security requirements are saved. Also, CRAW introduces storing a main list to manage mobile members' location while they move intra-group inter-area.*

## 1. INTRODUCTION

IP multicast is an efficient way for IP datagram delivery to multiple members. Its applications include electronic newspapers, multimedia file downloads, video conferencing and etc. Along with the development of wireless networks and also the increase of using mobile devices, the request for multicast applications such as mobile TV services, network games, and mobile commerce in wireless networks has grown up rapidly.

In both wired and wireless networks, the security should be guaranteed first. As IP multicast uses IGMP (Internet Group Management Protocol) [1] for group membership, multicast group is managed openly. This lack of security in multicast leads to eavesdropping. The solution for limiting access to multicast content is encryption with group key. A group key is the key which is shared by sender and all group members. Next, the security requirements of multicast are forward secrecy and backward secrecy [2-5].The former ensures that when a member leaves the group, it cannot access the current contents. The latter ensures when a member joins a group, it cannot access any past archive content. In order to provide these requirements, the group key needs to be updated on each membership change, join/leave, and to be delivered securely to valid multicast members. This process is known as group re-keying or re-keying in short.

A number of protocols have been proposed for group key management [6-14]. These protocols have two kinds of weaknesses. First of all, most of these protocols have been designed for wired networks and cannot be introduced to wireless networks because in addition to join/leave (Figure 1), in wireless networks, movement of members (Figure 2) implies more complexity to group key management. For example, in Figure 2, $u^4$ moves from area "A" to area "B" in the same group while $u^j$ joins multicast group in area "A" and $u^{12}$ leaves the same multicast group in area "B". Consequently, this additional behaviour, movement, increases the overhead of group key management largely in wireless networks.

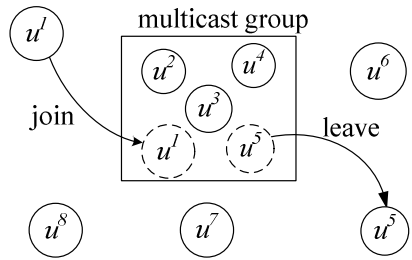

Figure 1.Join/leave in wired networks

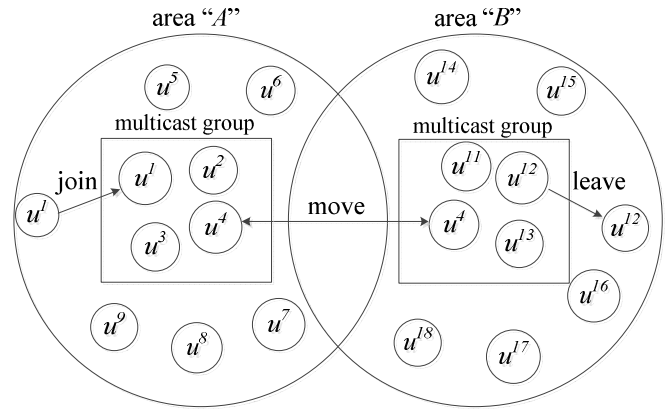

Figure 2.Join/leave and movement in wireless networks

The second obstacle of previously proposed protocols is separation of authentication and group key management. In fact, they have only focused on efficient key distribution to valid members. Because of features of wireless networks, join/leave and movement, the overhead of re-keying is larger than wired networks. In order to decrease it, one solution is to combine the authentication with an efficient group key management.

Some approaches address the group communication in wireless networks [15-17]. Without proposing any specific re-keying algorithm for mobile users in wireless environment, they have just suggested using one of the sophisticated methods proposed previously for group key management. However, in [18-20] new re-keying protocols for secure wireless multicast communications have been proposed. These protocols include distributed group key management method which is not able to reduce re-keying overhead for large number of mobile users who move between areas rapidly. Consequently, previously proposed protocols for mobile communication in wireless networks still suffer from re-keying overhead.

This paper proposes an efficient approach to decrease re-keying overhead in wireless networks. For this purpose, we combine authentication with group key management by using one-time password for authentication [21-23] and our previously proposed protocol [24] for group key management. Since the structure of wireless networks is very close to decentralized group key management, we use decentralized approach for our proposal. Finally, we add topology control to member movements in order to manage member location. For this purpose, a list of member movement information is stored in the main server.

This paper is organized as follows. Section 2 discusses the related work. Section 3 gives an overview to secure multicast, group key management and one-time password. The design principles and detailed design of our proposal are shown in section 4 and section 5, respectively. We discuss the security analysis of our protocol in section 6. We compare our approach with previously proposed group key management in section 7 and then complete our comparison with some simulation results. Section 8 is the conclusion.

## 2. RELATED WORK

As stated before, movement of mobile members between different areas complicates group key management in wireless networks. Several approaches have been proposed for secure multicast in wireless networks [15-20].The main purpose of these approaches is how to handle re-keying efficiently and securely while mobile members in intra-group inter-area join/leave the group. The design of most of them is based on decentralized network structure in which the network is divided into multiple areas, and each area is managed by an area controller responsible for generating the keys for the areas. Figure 3 illustrates the structure of such network where $AS_i$ is in *Area Server* (base station) for area $i$, $AK_i$ is an *Area Key* for area $i$, $M_x$ is a mobile member who moves between areas and the main server which is connected by wire to $AS.AK$ is generated for encrypting the group key as well as the group contents for area members and is sent to its area group members to be able to decrypt the group content.

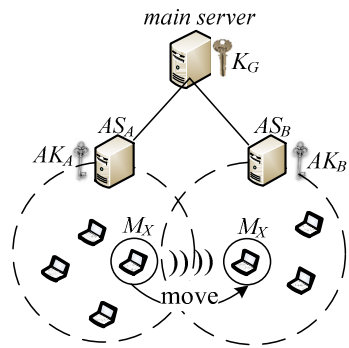

Figure 3.  Network partitions structure

The main server is located on top network level. In Figure 3, the network is divided into two areas. Each area has a base station for delivering the group content received from the main server. The users, who are under the supervision of the different area base stations, need to receive group services from them. In wireless networks, if a mobile member moves between areas, the area keys are updated to support the security requirements.

Static Re-key (SR) [15-17] proposes inter-area re-keying method. In SR (Figure 4(a)), the $AS$ maintains the keys unchanged when $M_x$ moves from area "A" to area "B".  So, movement of a member does not trigger any re-keying either in previous area or the new one. The $AS_A$ is still responsible for delivering its service when $M_x$ moves to "B" without any registration with the new local area server. Then obviously, mobility of $M_x$ affects neither the previous area re-keying nor the new area re-keying. Although this proposal has the least re-keying overhead, it suffers from the lack of security. Consequently, it cannot be supported with the most of applications which require security for content delivery.

Baseline Re-keying (BR) method [15, 16] tries to solve the mobility of group members as well as security. In this method, Figure 4 (b), movement of a member is considered as a leave of the group from the previous area and re-joins the group in the new area. For this purpose, group key and the area key are updated twice, in the previous area and in the new one. Consequently, the first drawback of BR is the interruption in service delivery through the old and new areas because of its high re-keying overhead. The second drawback is that the system is not able to recognize each join/leave of a member from his movement cross the areas. Moreover, these approaches do not propose any re-keying protocol for the group key management.

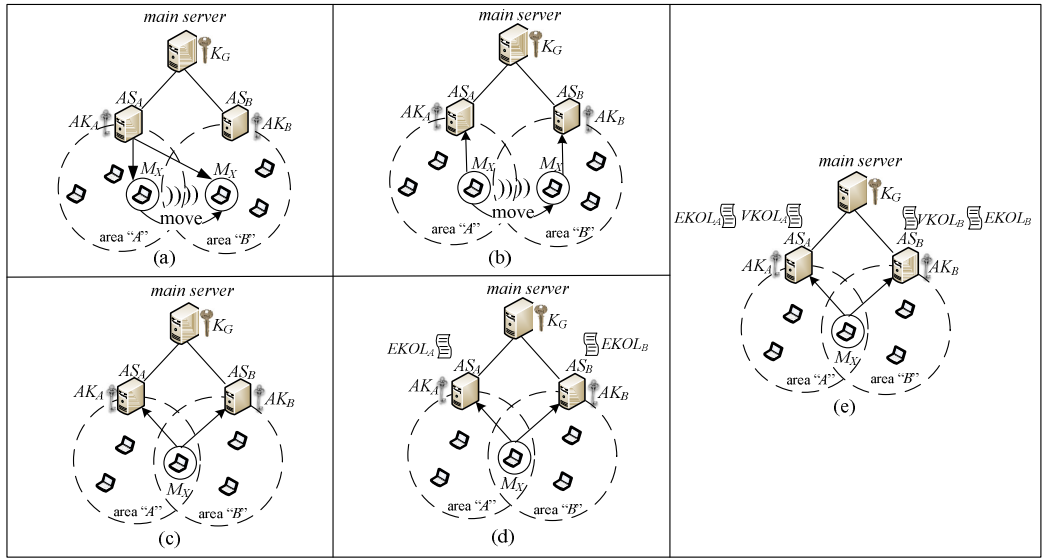

Figure 4.Area re-keying models (a) Static Re-keying (b) Baseline Re-keying (c) Immediate Re-keying (d) Pure Delayed Re-keying and FEDRP (e) Model of [17]

Immediate Re-key (IR) method [15, 16] is the extension of BR method. IR adds a hand-off mechanism to provide the member movement between areas. In this method (Figure 4(c)), when a member moves from area "A" to area "B", it sends a signalling message synchronously to both areas. Both area servers concurrently update area keys while the group key is remained unchanged. The features of IR are its continuous data transmission without interruption, and member hand-off mechanism. Despite of this fact, it still implies high re-keying overhead because no re-keying approach is proposed and using current group key managements are not compatible for applications which need rapid re-keying. A protocol with low re-keying mechanism is needed.

Delayed algorithm [15] proposes to postpone area re-keying until the pre-defined threshold is satisfied. The idea is based on the probability of returning the moving members to their previous area soon. Thus, the area re-keying in old areas is delayed for a threshold. That is, the old area key remains unchanged for reuse because that member may return promptly. Pure Delayed re-keying (Figure 4(d)) is similar to Delayed algorithm in the concept. In this method, local server maintains a list, *Extra Key Owner List (EKOL)*, from members who have moved to other areas. The key in old area is changed when a join/leave happens otherwise, the key is kept unchanged. *EKOL* is renewed whenever join/leave occurs in the area. By use of this list, when a member returns to its previous area, it is checked with the *EKOL* and if it is on that list, there will no need to update area key. The drawback of this approach is that backward secrecy is not assured till the next join/leave.

The other suggested method in this research area is Periodic re-keying [15].In this approach, area key update is done periodically without considering member movement. So, by periodically re-keying, no area key remains valid for more than a fixed pre-defined period of time. Actually, this proposal can be used in conjunction with other inter-area re-keying algorithms too.

First Entry Delayed Rekey+Periodic (FEDRP) [16], Fig. 4(d), is the combination of Delayed and Periodic re-keying protocols. When a member moves intra-group from area "$A$" to area "$B$", it sends concurrently two signalling messages to both area servers synchronously. Then, that member is added to *EKOL* in old area. *EKOL* is reset when the period is expired or a group join/leave is occurred. By this method, FEDRP improves significantly the inter-area re-keying by keeping old area keys unchanged. However, keeping *EKOL* seems to add additional storage overhead to previous area because it is more probable that a member enters a new area for the first time than is back to previous area.

Authors in [17] propose area re-keying in new visited area by introducing a specific key called *Visitor Encryption Key(VEK)*. *VEK* is delivered to a mobile member who moves to a new area. As shown in Figure 4(e), when a member moves form area "$A$" to area "$B$", it sends a signalling message to both area servers synchronously to notify about its mobility. Next, $AS_B$ sends the $VEK_B$ to $M_x$ using a secure channel. $VEK_B$ acts like $AK_B$ within the area. In this proposal, each area server stores two kinds of owner lists: *EKOL* as well as a *Visitor-Key Owner List*, *VKOL*, for mobile members. *VKOL* contains the list of members still holding a valid *VEK* but have left the area of that *VEK*. By this way, this method separates re-keying of mobile members from non-mobile members. As a result, no area re-keying is needed when a mobile group member moves between two areas. In fact, both backward and forward secrecies are assured along the mobility between areas. The drawback of this mechanism is that "one affects $n$ members". This means that each join/leave to the group causes re-keying for visitor members as well as area non-mobile members. Afterwards, separating the assigned keys of mobile members and the non-mobile ones is not a reasonable way for reducing the re-keying overhead.

Authors in [18, 19] propose a distributed group key management which divides group members into leaders and general members. The leaders participate in key management on behalf of general members. Authors believe that no area re-keying in both, old and new ones, is needed for member movement between two areas. For this purpose, hand-off member mechanism is defined to handle the member mobility. By this protocol, the backward and forward secrecy is not satisfied by maintaining the area key in both sides for mobile members. By using their distributed key management, they have reduced the re-keying overhead in comparison with LKH algorithm. However, the distributed architecture is not matched with wireless network structure. Therefore, it is not very useful in implementation.

Authors in [20] present a multicast key management tree for the mobile users in wireless networks. They design this tree according to wireless network topology. They claim that dividing the network into two separate levels, wired and wireless networks, improves the overall performance of re-keying in wireless networks.

# 3. OVERVIEW OF GROUP KEY MANAGEMENT PROTOCOL AND ONE-TIME PASSWORD

In this section, network structure of secure multicast is defined firstly. Then, we give a brief description of key tree structure. Next, an overview to our previously proposed group key management [22] is given. Finally, our suggested member authentication protocol [23] using one-time password is overviewed.

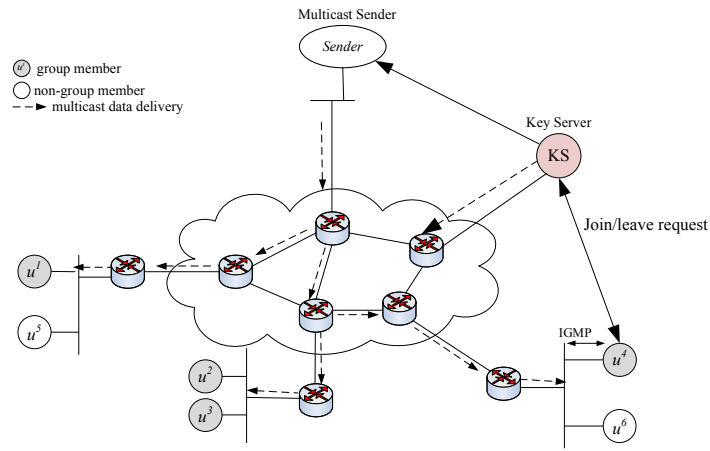

Figure 5. Network structure of group communications in secure multicast [21]

## 3.1. Network Structure

Figure 5 illustrates network structure of group communication in secure multicast. There exist three major entities, key server, multicast sender, and multicast receivers. The key server is responsible for generating and delivering the necessary keys to authorized group members. The multicast sender sends content to the group members. The group members are the users who are authorized for receiving the desired multicast services. On each group membership change, the necessary keys should be updated and distributed to group members securely. As illustrated in Figure 5, when a new member joins the group, $u^4$, it sends an IGMP request to its nearest router to receive multicast data from the network. Also, the new member, $u^4$, sends a join request to the key server. After being accepted, re-keying process is started. All the updated keys must be delivered to the multicast sender, new user, $u^4$, and all existing members, $u^1$, $u^2$ and $u^3$. When a member leaves the group, it sends an IGMP request to stop content delivery. Then, this member informs the key server by sending a leave message. At this time, re-keying for remaining members is performed.

## 3.2. An Overview to Structure of Logical Key Tree

In LKH based protocols [5-14], the members of multicast group are mapped with the leaves of a logical key tree (Figure 6). This tree is a d-ary tree, typically a binary one. Each member stores all the keys along the path from his leaf node to the root. The middle nodes are auxiliary keys which help the delivery of other keys. The root node is the group key. When a member joins or leaves the group, all the keys in his possession need to be changed to new ones. If the tree is binary, the new key is securely distributed by encrypting it using the two child keys. Since the height of the key tree is $log\ n$, the complexity of the key distribution is also $O(log\ n)$ (Figure 6).

Figure 6 shows a simple example of the logical key tree for a multicast group with seven members ($u^2$ through $u^8$). When $u^1$ joins the group, the node keys, $K_{1,2}$ and $K_{1,4}$, and the group key, $K_G$, are changed to $K'_{1,2}$, $K'_{1,4}$, and $K'_G$ respectively, because those nodes are on the path from $u^1$ to the root. Key generation, key encryption and key distribution for new member are done in the following order. First, all the keys which $u^1$ needs to have, $K'_{1,2}$, $K'_{1,4}$, and $K'_G$, are sent through unicast to $u^1$ being encrypted with $K_1$, $K'_{1,2}$, and $K'_{1,4}$ respectively.

For current members, the distribution is done in the following order. $K'_{1,2}$ is sent to $u^2$ by unicast, being encrypted with $K_2$.$K'_{1,4}$ is sent by multicast to $u^2$ and to {$u^3$,$u^4$} being encrypted with $K'_{1,2}$ and $K_{3,4}$, respectively. Also, $K'_G$ is sent by multicast for {$u^2$, $u^3$, $u^4$} and for {$u^5$, $u^6$, $u^7$, $u^8$}, being encrypted with $K'_{1,4}$ and $K_{5,8}$, respectively. When a member leaves the group, key generation, key encryption and key delivery to remaining members is done in the same manner.

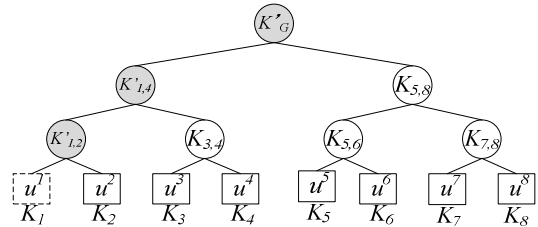

Figure 6. The logical key tree for LKH based approaches at join/leave [22]

## 3.3. Code for Key Calculation (CKC)

CKC [22] is the most efficient group key management protocol which can be used in wireless networks because it focuses on user side key calculation rather than server side key distribution. CKC inherits the concept of logical key tree from LKH. In CKC, when a new member joins the group, server sends only the group key for that new member. Then, current members and that new member calculate the necessary keys by node code and one-way hash function. Moreover, at leave, only the group key is sent to remaining members, then those members calculate the necessary key by using node code and one-way hash function.

A node code is a random number which is assigned to each node to help users calculate necessary keys. A new node code is generated by concatenating a random number to right digit of its parent node code, *child_node_code= (parent_node_code|| random_digit)*. Figure 7 shows the node code assignment in key tree. For example, when a node code is *645* and the generated random number is *2*, the code assigned to that node will be *6452*. Moreover, when the number of nodes increases, for each new node, the same process is repeated. Generally, each member can reach to his parent node code by deleting the right digit from the child code. By this mechanism, each member knows the codes of all nodes in the path to the root. So, members update affected node keys by these codes and applying hash function after each membership change.

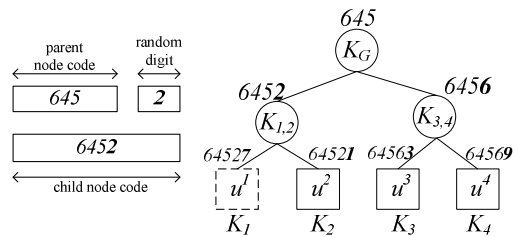

Figure 7. Node code management in CKC

In CKC, when a new member joins the group, it is allocated to a node in key tree. Then, the node code and individual key are delivered to it securely. The server encrypts the group key with the new member's individual key and sends it to that member by unicast. The current members apply a one-way hash function to previous group key to calculate the new one. When f is a one-way hash function and $K_G$ is the previous group key, the new group key, $K'_G$, is calculated through the formula below by current members.

$$K'_G = f(K_G)$$

The affected middle node keys are updated by this formula in user side. Where $C_{middle\_node}$ denotes the new code of middle node and $K'_G$ is the current group key, and $K'_{middle\_node}$ is the middle node key in the path to the root.

$$K'_{middle\_node} = f(K'_G \oplus C_{middle\_node})$$

When a member leaves a multicast group, the node of leaving member is deleted from the key tree, and its sibling member is moved to its parent position node. Then, the key server generates a new group key and then encrypts it by the keys of top nodes on each half of the key tree. Again, related middle node keys are computed by members themselves using node codes and one-way hash function.

Obviously, CKC reduces key generation and key encryption overhead largely at join/leave compared to other LKH based protocols. Moreover, the unicast and multicast communication overhead are decreased at join by CKC method largely. In fact, these properties are needed for secure multicast in wireless networks. Because, in such networks, in addition to join/leave, the inter-area intra-group movement of a mobile member is also considered as leaving the old area and joining new area.

### 3.4. One-Time Password Protocol

One-time password protocol SAS (Simple and Secure) authentication protocol, [23, 24], consists of two phases, registration and authentication. At the registration phase, the supplicant registers a hash value which is generated by bitwise XOR of its password, $S$, and a random number, $N_i$, to the server. After a period of time when the authentication is needed, it moves to authentication phase. At this phase, the registered member generates two other hash values, $\alpha$ and $\beta$, generated by bitwise XOR of $S$ and a new random number, $N_{i+1}$, and sends to the server. Then, the server performs some calculations on $\alpha$ and $\beta$ to verify if the extracted value matches the registered one. When the values match, the user is authenticated. Part of this extracted hash value at the authentication phase is registered for re-authentication. The details of this algorithm will be explained by an example in section 5.

## 4. DESIGN PRINCIPLES

We introduce our efficient re-keying approach for mobile members in wireless networks. The purpose of this protocol is to reduce re-keying overhead without any interrupt in content distribution while security requirements are assured. For reducing re-keying overhead, we use our previously proposed group key management protocol, (CKC) and combine it with authentication operation. In addition, we introduce members' topology control list to manage mobile user location in wireless networks. The following steps show the design principles of our approach.

Figure 8 illustrates network structure of secure multicast in wireless networks. This network has four elements; the main server, *Area Wireless Server* (*AWS*), mobile members and a main list. The main server is responsible for sending group content to each *AWS* and storing the main list. The main list contains mobile members' information regarding join/leave and movement. Each *AWS* is responsible for authentication, generating and sending group key, and finally transmitting content to mobile members. The main server has wired connection to all area wireless servers through the network. In this level, main server and *AWS*s share a specific key. As this figure shows, wireless network is divided into several areas. Each *AWS* is responsible for connecting mobile users to the network. In wireless network level, mobile members move between areas while remaining in the same group.

The main list stored in the main server supports users' topology control and user authentication information. Authentication information of registered members is saved in this list globally. Different *AWS*s refer to this list to recognize authorized group members' location or their join/leave information. The information in the main list enables servers to have topology control on different mobile group because all the information about members' behaviour, join/leave and movement are necessary to support applications which need accounting.

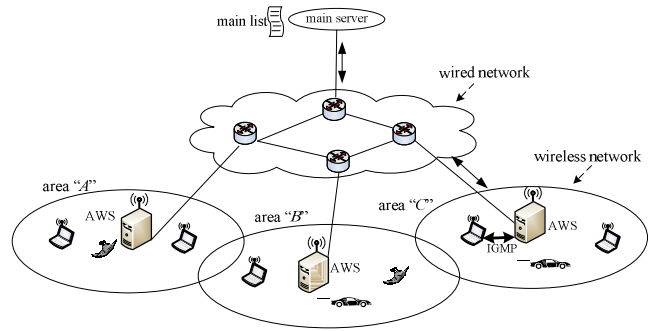

Figure 8.Network structure of secure multicast in wireless networks

In this step, we discuss about join/leave process of mobile members through different areas of wireless networks. As mentioned before, mobile users in each area are managed by the wireless server of that area. Different *AWS*s support different collections of multicast groups. The process of join/leave is as following steps.

(1) When a mobile user joins a multicast group, it sends IGMP message to its *AWS*.

(2) Then, *AWS* forwards this message to the nearest router to connect to the main server.

(3) The new user sends join request to the *AWS* for receiving *Area Group Key* ($AK_G$) of the group which wants to join.

(4) Before delivering the $AK_G$, the *AWS* performs user authentication procedure using one-time password protocol.

(5) If user authentication is done successfully, the $AK_G$ is updated and is sent to that new user by *AWS*.

(6) At this time, main server sends the desired group content to the *AWS*.

(7) Next, *AWS* delivers the group service to the new member. The remaining group members in this area generate the new $AK_G$ themselves using CKC protocol.

(8) When a member leaves the group, it sends a leave request to its *AWS*.

(9) Next, the *AWS* forwards leave request to main server for stopping its content delivery. *AWS* also registers the information of that leaving member to the main server to add it to the main list. Next time when this member re-joins, it is authenticated by this information.

In this part, the movement of mobile members between areas is explained. During each member movement, $AK_G$ of both areas, the old and new one, are updated to provide forward and backward secrecy respectively. We use a hand-off mechanism for handling the member movement. Figure 9 illustrates mobile member movement policy, when that member moves from its current area to new one. The movement process is described as below.

(1) Mobile member sends two synchronous requests to the old *AWS* and the new one.

(2) The old *AWS* sends the leaving member's information to main server to store this information in the main list.

(3) Before accepting member in the new area, the new *AWS* requests this member's information from the main server.

(4) By receiving the response from the main server, the new *AWS* starts to authenticate the new joining member.

(5) If the authentication is successful, new $AK_G$ of the new area is delivered by new *AWS*. Until receiving new $AK_G$, the old *AWS* is responsible for delivering group content to the moving member.

(6) Synchronously, when new *AWS* delivers the new $AK_G$ to that mobile member, it sends an area join acknowledgment to the previous *AWS* for stopping the content distribution.

(7) The old *AWS* updates its $AK_G$ to ensure backward secrecy.

(8) At the end, new *AWS* sends the group content to the new area member.

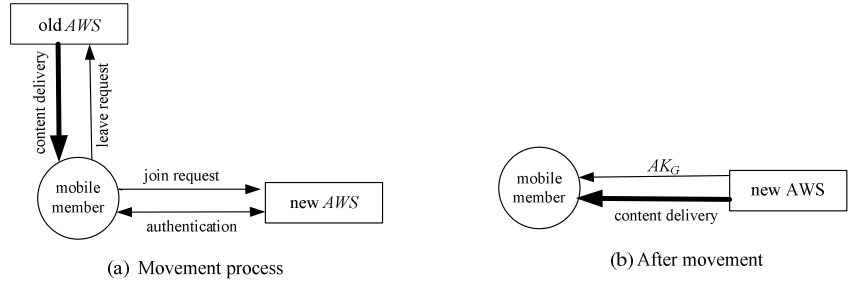

Figure 9.Movement policy

Finally, our work has some useful features which distinct it from the other approaches. Simplicity is the most important feature of our approach. We combine authentication with group key management to use authentication information instead of member individual key. To increase efficiency of this approach, we propose using a group key management protocol with low overhead for area re-keying. Unlike the other approaches, our approach provides forward and backward secrecy together while member movement is occurred.

## 5. DETAILED DESIGN

This section presents the details of our approach, CRAW (Combination of Re-keying and Authentication in Wireless Networks).

### 5.1. Notation

Before we proceed further, we introduce the notations which we have used for explanation of our approach (Table 1).

Table 1. Notation

| | |
|---|---|
| $S$ | user password |
| $i$ | number of authentication session |
| $N_i$ | random number corresponding to $i^{th}$ authentication |
| $E$ | hash function for authentication |
| $E(x)$ | $x$ is hashed once |
| $E_2(x)$ | $x$ is hashed twice |
| $j$ | member index |
| $E^j(x)$ | hash value registered for member $j$ |
| $\oplus$ | bitwise XOR operation |
| $K_{m,n}$ | node key with depth of $m$ and width of $n$ |
| $u^j$ | member $j$ |
| $K_j$ | key of user $j$ |
| $AK_G$ | group key shared in the group through an area |
| $f$ | one-way hash function for group key management |

### 5.2. Member Join Operation

We now use Figure 10 and Figure 11 to explain how authentication and join operation for a multicast group is performed in wireless networks. For simple explanation, we consider a group of 7 members, $\{u^1, u^2, u^3, u^4, u^5, u^6, u^7\}$, when $u^8$ joins the group in area "B". The $AWS_B$ is the manager of area "B". The following steps show the join process of $u^8$ in details.

(1) For connecting to the network, $u^8$ sends IGMP message to $AWS_B$ to be able to receive the multicast group content.

(2) After that, $u^8$ sends a join request to $AWS_B$ to join the group. $AWS_B$ starts to authenticate $u^8$ with the main server. The user authentication is performed by two phases of one-time password protocol.

*I) Registration phase:*

We assume that $u^8$ is a registered member and its information has been stored in the main list. By this assumption, when $N_1$ is a random number and $S$ is the password of $u^8$, the registered information of $u^8$ is $E^8(N_1 \oplus S)$.

*II) Authentication phase:*

- $u^8$ generates another random number, $N_2$, and performs bitwise XOR with $S$, and then finds $E^8(N_2 \oplus S)$. Here, $u^8$ calculates the following two other hash values, $\alpha$ and $\beta$, using $E^8(N_2 \oplus S)$ and sends them to $AWS_B$.

$$\alpha = E_2^8(N_2 \oplus S) \oplus E^8(N_1 \oplus S)$$
$$\beta = E^8(N_2 \oplus S) \oplus E^8(N_1 \oplus S)$$
$$u^8 \xrightarrow{\alpha, \beta} AWS_B$$

- $AWS_B$ uses $\beta$ and the previously registered information of $u^8$, $E^8(N_1 \oplus S)$ and extracts $E^8(N_2 \oplus S)$.

$$\beta \oplus E^8(N_1 \oplus S) = E^8(N_2 \oplus S)$$

- Then, $AWS_B$ operates one more hash function on $E^8(N_2 \oplus S)$ and finds $E_2^8(N_2 \oplus S)$. It performs again bitwise XOR on $\alpha$ and $E_2^8(N_2 \oplus S)$.

$$\alpha \oplus E_2^8(N_2 \oplus S) = E^8(N_1 \oplus S)$$

- At the end, $AWS_B$ compares the result of the above calculation, $E^8(N_1 \oplus S)$ with the previously registered data of $u^8$ in the main list to verify if they match. If two values are equal, authentication succeeds. $E^8(N_1 \oplus S)$ is used for individual key.

(3) Now, $AWS_B$ accepts $u^8$ as one of its new area members. Re-keying process in area "B" is done as below.

I) $AWS_B$ updates the area group key from $AK_G$ to a new one, $AK'_G$ using one-way hash function.

$$AK'_G = f(AK_G)$$

II) $AWS_B$ sends $AK'_G$ and the position code by unicast to $u^8$ being encrypted by $K_8$.

$$AWS_B \xrightarrow{unicast} (AK'_G, 1578)_{K_8}$$
$$K_8 = E^8(N_1 \oplus S)$$

III) The current group members,$\{u^1, u^2, u^3, u^4, u^5, u^6, u^7\}$,compute $AK'_G$ by using one-way hash function of previous group key by themselves.

$$u^1, ..., u^7 : AK'_G = f(AK_G)$$

IV) The parent node key of the new node, $K_{5,8}$, is updated to $K'_{5,8}$ by the users according fallowing formula. And also the new node, $K_{7,8}$, is generated as a parent node for $u^7$ and $u^8$. For this purpose, they use the node codes and the new area group key.

$$u^5, ..., u^8 : K'_{5,8} = f(AK'_G \oplus 157)$$
$$u^7, u^8 : K_{7,8} = f(AK'_G \oplus 1578)$$

V) Then, the main server sendsthe requested content for $AWS_B$ through the wired and$AWS_B$ starts to deliver the content to the group members in area "B".

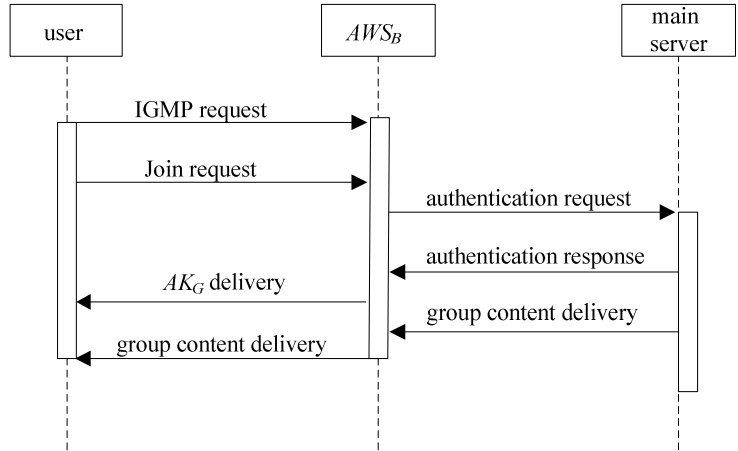

Figure 10.Join process in wireless networks

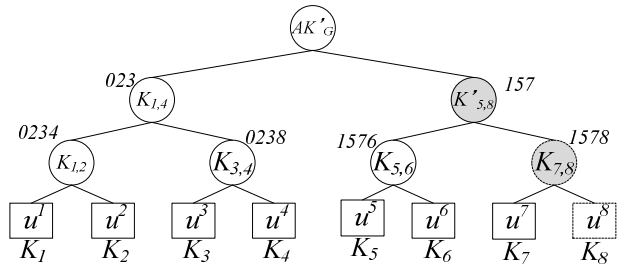

Figure 11.The key tree in $AWS_B$ when $u^8$ joins the group

## 5.3. Member Leave Operation

For simplicity to explain leave operation we use a simple example of 8 members in area "A" when $u^8$ leaves the group. Figure 12 illustrates the leave procedure in area "A" and Figure 13 illustrates the key tree in area "A" after the leave of $u^8$.

(1)  $u^8$ sends leave request to $AWS_A$.

(2)  $AWS_A$ forwards the leave request to the main server for stopping its content delivery.

(3)  For the re-join purpose, $AWS_A$ sends the information of $u^8$to the main server to be stored in the main list. $AWS_A$ registers $E^8(N_2 \oplus S)$ for re-authentication of $u^8$at re-join.

(4)  To provide forward secrecy, when $u^8$leaves the group, another re-keying process is needed. Again, the node of $K_{7,8}$ is deleted and $u^7$ is promoted to its position. New $AK_G$ is generated by $AWS_A$; this new area group key is encrypted by top node keys of each part, and then is sent to remaining members of each part by multicast.

$$AWS_A \xrightarrow{multicast} \begin{cases} u^1,...,u^4 : (AK'_G)_{K_{1,4}} \\ u^5, u^6 : (AK'_G)_{K_{5,6}} \\ u^7 : (AK'_G)_{K_7} \end{cases}$$

(5) Then, the remaining members in the path to the root,$\{u^5,\ u^6,\ u^7\}$, use $AK'_G$ to update the affected middle node keys as following.

$$u^5,\ u^6, u^7 : K'_{5,7} = f\left(AK'_G \oplus 762\right)$$

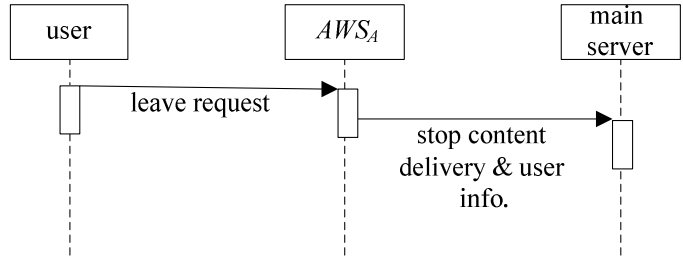

Figure 12. Leave policy

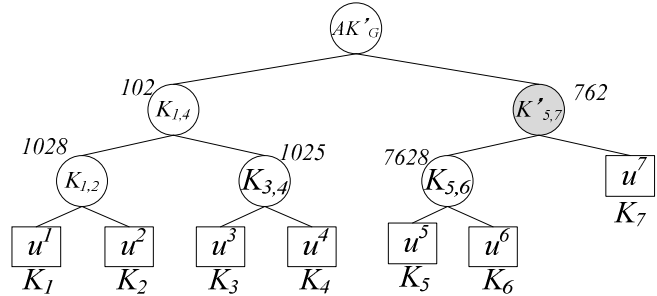

Figure 13. The key tree in $AWS_A$ when $u^8$ leaves the group

## 5.4. Member Movement Operation

We assume that $u^8$ has leaves the group from the area "$A$" and moves to area "$B$", while remaining in the same group. Figure 14 and Figure 15 explain the movement procedure when $u^8$ moves from area "$A$" to area "$B$".

(1) When $u^8$ reaches to the edge of area "$B$", it enters to hand-off mechanism. $u^8$ sends a hand-off leave request to $AWS_A$ and synchronously sends a hand-off join request to $AWS_B$. Besides that, $u^8$ sends authentication request which contains two hash values ($\alpha$, $\beta$) to $AWS_B$ too.

(2) At this time, $AWS_A$ delivers the information of $u^8$ to the main server for storing in the main list. This user information contains data about the member's group name, type and amount of its received service and the last location of that member.

(3) $AWS_B$ starts to authenticate new member by sending an authentication request to the main server about $u^8$.

(4) The main server looks for registered authentication information of $u^8$ in its main list and sends it to $AWS_B$ for re-authenticate. We assume that the available registered authentication information is $E^8\left(N_1 \oplus S\right)$.

(5) Now, $AWS_B$ moves to authentication phase of $u^8$ and authenticates $u^8$ as explained in subsection 5.2.

(6) At the end of authentication process, re-keying is performed in area "*B*" as explained in subsection 5.2.

(7) At this moment, $AWS_B$ sends an area join acknowledgement for $AWS_A$ to stop delivering the service to $u^8$. By this signalling, $AWS_A$ does re-keying process in its area for supporting forward secrecy as explained in subsection 5.3.

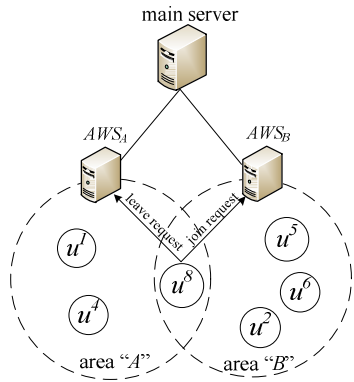

Figure 14.Hand-off mechanism

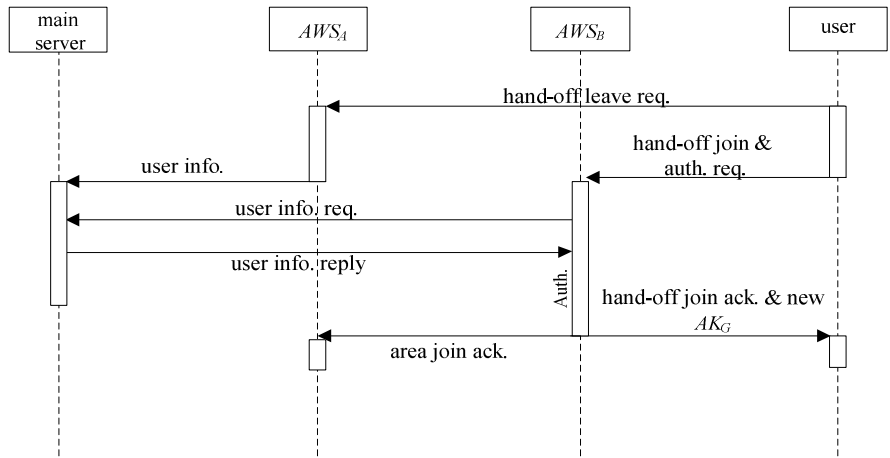

Figure 15.Member movement in wireless networks

# 6. DESIGN FEATURES AND REQUIREMENTS

In this section, we analyse and discuss design features, requirements, and compare CRAW with other conventional methods logically. Starting from authentication, CRAW suggests one-time password for authentication. Using one-time password for authentication has two advantages. Firstly, a part of user authentication information is stored in the server, and each time the user is authenticated by a new password. If the password information is eavesdropped in a session, only the information of that session is lost. Secondly, user authentication information can be used as an individual key for the group member. In fact, the property of one-time password which is refreshed on each session at authentication provides this opportunity to be used as member individual key. Consequently, the requirement for member individual key and the property of one-time password for member authentication are completely fit. Because of this feature of CRAW, at join or inter-area movement of member, the *AWS* does not need to generate individual key for the member. Consequently, this feature decreases the overhead of group key management in wireless networks when movement of the members in new areas dynamically happens.

Next, CRAW is a simple method in design compared with previously proposed ones. Other traditional methods sacrifice security requirements of secure multicast and complicate re-keying process to decrease the re-keying overhead with different ways. In compare, the essence of CRAW is simple; member authentication, join/leave and movement. In CRAW, the movement of a member to an area is considered as a leave from the old area and join to a new area. This is because of security requirements; forward secrecy in old area and backward secrecy in new area. Moreover, when a member moves inter-area, individual key needs to be delivered to that member, as in CRAW authentication information is used for individual key as well, the server does not need to generate individual key and deliver it to that member. In addition, CRAW takes advantages of CKC for group key management because in CKC members participate to generate the necessary keys rather than delivered by *AWS*. In fact, when a member joins a group or moves to a new area, *AWS* only delivers *AKG* to that member.

In addition, keeping a main list in the main server is useful to manage movement of the mobile members. In wireless networks, it is required for *AWS*s to recognize users' location and their latest status. This is crucial for multicast applications which need member authorization for accounting. This important requirement is not seen in design of conventional approaches.In compare, CRAW has this ability which includes user authentication information, user location information, and the last, group membership.

## 7. COMPARISON AND RESULTS

In this section, we analyse the efficiency of CRAW in details. First, we compare computational and communication overhead of our suggested group key management, CKC, with the most typical group key management protocol, LKH. In our comparisons, *n* denotes the group size which is the number of members in an area after join and before leave operations. The performance of these protocols depends on the number of key generations, key encryptions, and key transmissions for members at join/leave.Tables 2, 3 and 4 summarize our comparisons, focusing on the following measures:

- Computational overhead
  - ✓ Key generation overhead: The number of keys that must be generated at join/leave.
  - ✓ Encryption overhead: the number of encryptions.
- Communication overhead: the number of transmissions from the key server.

Table 2 shows the key generation overhead at join/leave operation. In LKH group members do not generate any keys in each join/leave operation. Therefore, in this protocol,the server has the most loads for key generation which is equal to tree depth and the key generation overhead is$\log_2 n$. In CKC, the number of key generations at join operation is *2* and at leave is *1*. It is because at join, the new user individual key and group key, and at leave, only new group key need to be generated by the server. Moreover, when this protocol is combined with authentication the number of key generation at join is decreased to*1*. This mechanism reduces key generation overhead in CKC.

Table 2.The comparison of key generation overhead at join/leave operation

| Protocols | Join | Leave |
|-----------|------|-------|
| LKH | $\log_2 n$ | $\log_2 n - 1$ |
| CKC | 2 | 1 |

Table 3 shows the key encryption overhead at join/leave operation. The number of key encryptions for LKH is $3\log_2 n$ which is large. It is because the server needs to encrypt all the updated keys with their children keys. The key encryption overhead in CKC is $1$ which is small at join. It is because server encrypts only the new group key for the new user. CKC reduces the number of key encryptions at join, and keeps it at leave at the lowest level. At this time, server encrypts new group key by top node of each part for users of that part. So, the key encryption overhead in CKC at leave is $\log_2 n$. The purpose of design of CKC is to reduce computational overhead at join for wireless networks because in such networks simultaneous join/movement of members to a new area which causes large re-keying overhead in new area is more probable. So it is required to reduce the overhead of join operation.

Table 3. The comparison of key encryption overhead in join/leave operation

| Protocols | Join | Leave |
|-----------|------|-------|
| LKH | $3\log_2 n$ | $2\log_2 n$ |
| CKC | 1 | $\log_2 n$ |

Table 4 illustrates the communication overhead at join/leave for LKH, and CKC. Communication overhead at join is divided into two categories, unicast overhead for the new member and multicast overhead for the current members. As shown in Table 4, unicast overhead for CKC is equal to $1$ because only the group key is delivered to the new user. But the unicast overhead is equal to the height of the key tree for LKH. The multicast overhead at join for LKH is equal to height of the key tree while CKC has no multicast overhead because no key is delivered to the current members at join. On the other hand, when a member leaves the group, the communication overhead for CKC is equal to the height of the key tree while LKH has two times of transmission message numbers in compare with CKC transmissions.

Table 4. The comparison of communication overhead at join/leave operation

| Protocols | Join | | Leave |
|-----------|------|------|-------|
| | unicast | multicast | multicast |
| LKH | $\log_2 n$ | $\log_2 n$ | $2\log_2 n$ |
| CKC | 1 | - | $\log_2 n$ |

Table 5 illustrates summary of re-keying cost and the required time for joining a new member. As the previous methods for group key management are based on LKH, the re-keying cost of them at join is $\log_2 n+1$. CKC reduces this cost largely to $2$. But in CRAW, by using one-time password protocol for member authentication, the re-keying cost of CKC has been decreased to $1$, because the individual key is equal to user's authentication information in CRAW. At leave operation, our method has the same re-keying cost, $\log_2 n$ which is similar to the previously proposed ones. Another important result is about the required time when a new member joins the group, $\beta$, which is smaller than join required time in previous method, $\alpha$. In fact, using one-time password protocol for member authentication decreases the authentication time.

Table 5. Summarized of re-keying cost and member authentication time

| Methods | Re-keying Cost | | Required Join Setup Time |
|---------|------|-------|--------------------------|
| | join | leave | |
| Previous | $\log_2 n+1$ | $\log_2 n$ | $a$ |
| CRAW | 1 | $\log_2 n$ | $\beta\ (\beta < \alpha)$ |

When a new member sends a join request to the server, member authentication and preparing a unique individual key for that member are performed respectively. We call this process join setup time. At the end of join-setup process, the member's individual key is delivered through a secure channel.We compare join-setup time of CRAW with other models for wireless networks.$T_{ordinary-join-setup}$ and $T_{CRAW-join-setup}$ are defined as the time required for join-setup process in the other modelsand CRAW respectively. According to the following formula, the ordinary join-setup time contains two time intervals which belong to the preparation of individual key for new member and also the authentication process. However, the CRAW join-setup time just includes the member authentication period.

$$T_{ordinary-join-setup} = T_{authentication} + T_{individual\,key\,preparation}$$

$$T_{CRAW-join-setup} = T_{authentication}$$

Note that in the individual key preparation include two different operations, key generation and key distribution. Here, we do not consider the key encryption mechanism because the individual key is sent for the new member through the secure channel. The following formula shows the time needed for individual key preparation.

$$T_{individual\,key\,preparation} = T_{key\,generation} + T_{key\,distribution}$$

In CRAW, we use one-time password protocol for member authentication. By using one-time password method, we do not have any time consuming for individual key generation and distribution for a new member. The other approaches do not consider a specific authentication algorithm and they have just assumed that member authentication is occurred by a regular authentication protocol. Although the authentication time with one-time password is a little bit longer than general authentication methods, the overhead time gets shorter than the individual key preparation time in other approaches. Consequently, CRAW join-setup time is smaller than the other's. Because the difference of join-setup time of CRAW and the ordinary ones, $\delta$, is less than the total needed time for an individual key generation and distribution.

$$T_{ordinary-join-setup} > T_{CRAW-join-setup}$$

$$\begin{cases} T_{ordinary-join-setup} - T_{CRAW-join-setup} = \delta \\ \delta = T_{key\,generation} + T_{key\,distribution} \end{cases}$$

To show the correctness of the above equations, we have performed an experiment to collect the average time of member authentication with one-time password protocol and the traditional authentication algorithm. The simulation source code is in PHP language. The simulation has been run on two parts, the client side and the server side. The hardware information of the machine is 2.83 GHz, Intel(R) Core(TM) 2 Quad with CPU 708.351 MHz processor speed. The total memory is and 524288 in the server side. On the other hand, the information of the machine in the client side is a 2.67 GHz Windows Vista processor with 4 GB of RAM.In this experiment, we have collected the authentication time of *100* users. Table 6 summarizes our results for member authentication time.

Table 6.The comparison of member authentication time with different protocols

| Authentication Protocol | Minimum (sec) | Maximum (sec) | Average (sec) |
|---|---|---|---|
| One-time Password | *0.000669* | *0.005117* | *0.002517* |
| General | *0.000123* | *0.00086* | *0.000237* |

To find the individual key generation and distribution time, we have used a program to measure $T_{individual\,key\,preparation}$ of the conventional methods. This program computes the processing time of generating and delivering a random individual key. Our source code is based on the script language program, ruby, with OpenSSL cryptography library. We run this program on a 2.00 GHz Windows Vista processor with 2 GB of RAM. The results show that the required time forgenerating and distributing an individual key for a new member is *0.939000 (sec)*.According to the achieved results from Table 6 and using the above equation, the join-setup time for CRAW and the conventional methods is as below.

$$T_{ordinary-join-setup} = 0.000237 + 0.939000$$
$$= 0.939237 _{(sec).}$$
$$T_{CRAW-join-setup} = 0.002517 _{(sec).}$$
$$T_{ordinary-join-setup} - T_{CRAW-join-setup} = 0.93672 _{(sec)}$$

As mentioned in section 5.4, another feature of CRAW is hand-off mechanism for a mobile member who moves inter-area intra group. We implement our policies for member movement through a simple network with a simulator program, OMNeT++, version 4.1b4. Figure 16 illustrates our graphical result on network delay which is caused by our hand-off mechanism during the simulation time. In our experiment, a mobile member leaves the current area, and then joins a new area.

In fact, three operations occur in the joining area respectively; probe, re-authentication and re-association process. First, when a mobile member receives to the edge of its area, it starts to search for a new *AWS* to connect it. In this duration, the mobile member sends some probe requests to different *AWS*s until valid *AWS* responses. Second, when the moving member sends its join request to the valid *AWS*, re-authentication phase is started. The new *AWS* tries to re-authenticate the mobile member with the help of the main server. After accepting the mobile node's join request by the new *AWS*, re-association procedure is started. In this period, *AWS* performs group key management in its area, and then sends necessary keys for the new member and current area members. Obviously, each of these three processes causes some delay period in content delivery to the mobile member. According to the following formula, the sum of these delays is the total delay which hand-off mechanism implies to the network.

$$T_{CRAW-hand-off} = T_{probe} + T_{re-authentication} + T_{re-association}$$

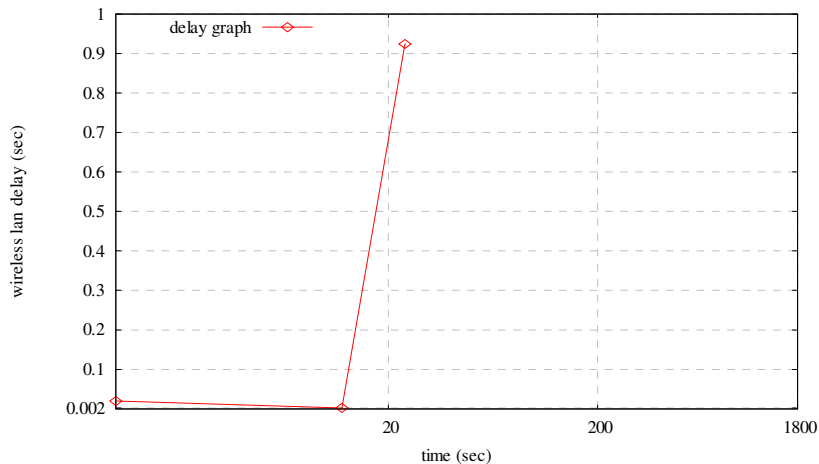

Figure 16.Hand-off delay for a mobile member

Figure 16 shows wireless delay time (sec) versus our simulation time (sec) for the moving mobile member. Three delay intervals are occurred in delay graph of this mobile member. The first delay, 0.0195167 (sec), is the time which is spent for finding a valid *AWS*. The second delay is 0.002517 (sec) which is the re-authentication time that the new *AWS* causes this delay. At last, after probe and re-authentication delay, new *AWS* does re-keying and sends the new area group key for the new user. This task causes 0.924 (sec)delay. After joining the mobile user to the new area, we do not have any delay until the end of the simulation because it remains in this area, and just receives the group content. At the end of the simulation, we calculate CRAW hand-off time which is equal to sum of all delays of the mobile node, 0.9460337 (sec). Our numerical values of hand-off time parameters are illustrated in Table 7.

Table 7.Hand-off delay time parameters

| Delay Parameters | Time (sec) |
|------------------|------------|
| Probe | *0.01951670* |
| Re-authentication | *0.00251700* |
| Re-association | *0.92400000* |

According to Table 7, hand-off re-authentication time of CRAW is equal to 0.002517 which is the same as authentication time with one-time password protocol (Table 6). On the other hand, CRAW uses CKC for re-association in each wireless area, therefore the lowest computational and communication overhead is occurred. By using CKC in re-association, just the new area group key is sent to the new area member. This re-keying mechanism takes a small period of time in compare with the other protocols. As stated before, the other area re-keying methods use the LKH based protocol for re-association. Therefore, they send new area group key to the new area member along with sending it to the current area members. In other words, the conventional models have much more re-keying overheads; therefore they cannot have an efficient re-association time.

Consequently, the other area re-keying methods do not propose directly a group key management protocol. They try to suggest other mechanisms, not an efficient re-keying algorithm, for reducing the re-keying cost through mobile member movement process. Obviously, it is not a correct comparison if we want to compare CRAW hand-off mechanism with them, because they do not describe their hand-off method. On the other hand, none of the previously proposedmethods measure their hand-off time exactly. In the previously proposed methods, they claim that using their methods can reduce re-keying overhead when a movement is occurred. Actually, in most cases, they suggest to reduce number of key update operation in network areas. Finally, the other methods just have focused on the number of re-keying procedures while we focus on using an efficient group key management protocol, and in addition having a secure and efficient authentication protocol in CRAW method. We examined the efficiency of CRAW based on its details and duration time of each step along with the logical proof.

## 8. CONCLUSIONS

We conclude our approach with its some contributions and features. In this paper, we have combined the group key management with the user authentication in wireless networks for the first time. In CRAW, re-keying process is done by CKC, and the multicast member authentication is performed by one-time password protocol. Unlike the other methods, CRAW provides a simple and secure mechanism for supporting both forward and backward secrecy with the lowest re-keying overhead. At the end, we summarize our result with some of its contributions and features:

- CRAW achieves high performance by separating wired and wireless networks for group key management.
- CRAW reduces re-keying overhead largely in each join/leave and movement.
- CRAW supports secure inter-area movement for mobile member.
- CRAW uses authentication information for group key management in a simple and secure method.
- CRAW benefits from the speed of hash function for authentication mechanism.
- CRAW is able to control the user location through the networks by saving a central list.
- CRAW supports accounting applications using member topology control.
- In CRAW, the group content distribution is performed continuously without any interruption.

**Authors**

**Elina Eidkhani**received her B.Sc.and M.Sc. degrees in Information Technology fromSharif University of Technology, International Campus, Iran. Her research intrests include security, wireless networks, IP Multicast, Group Key Management Protocols in various type of networks.

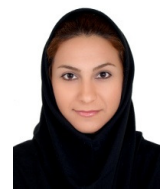

**Melisa Hajyvahabzadeh**received her B.Sc. and M.Sc. degrees in Information Technology fromSharif University of Technology, International Campus, Iran. Her research intrests include security, IP Multicast, Group Key Management Protocols, authentication in various type of networks.

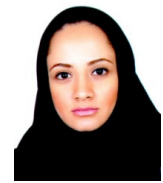

**Seyedeh Anahita Mortazavi** received her B.Sc.andM.Sc. degrees in Information Technology fromSharif University of Technology, International Campus, Iran. Her research intrests include security, IP Multicast, Group Key Management Protocols in various type of networks.

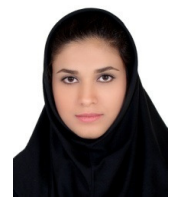

**Alireza Nemaney Pour** has obtained his B.S degree in Computer Science from Sanno University, Japan, M.S in Computer Science from Japan Advanced Institute of Science And Technology, Japan, and Ph.D. degree in Information Network Science from Graduate School of Information Systems,the University of Electro-Communications, Japan.

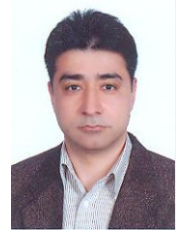

He is currently a faculty memberof Islamic Azad University of Abhar in Iran. In addition, He is a technical advisor of J-Tech Corporation in Japan. His research interests include Network Security, Group Communication Security, Protocol Security, Information Leakage, Spam Mail Prevention, Web Spam Detection, Group Authentication, and Cryptography.